**Exciton-Photonics: From Fundamental Science to Applications**

**Surendra B. Anantharaman, Kiyoung Jo, and Deep Jariwala\***

Electrical and Systems Engineering, University of Pennsylvania, Philadelphia, PA, 19104, USA

**\*Email:** dmj@seas.upenn.edu

## Abstract

Semiconductors in all dimensionalities ranging from 0D quantum dots and molecules to 3D bulk crystals support bound electron-hole pair quasiparticles termed as excitons. Over the past two decades, the emergence of a variety of low-dimensional semiconductors that support excitons combined with advances in nano-optics and photonics has burgeoned a new area of research that focuses on engineering, imaging, and modulating coupling between excitons and photons, resulting in the formation of hybrid-quasiparticles termed exciton-polaritons. This new area has the potential to bring about a paradigm shift in quantum optics, as well as classical optoelectronic devices. Here, we present a review on the coupling of light in excitonic semiconductors and investigation of the unique properties of these hybrid quasiparticles via both far-field and near-field imaging and spectroscopy techniques. Special emphasis is laid on recent advances with critical evaluation of the bottlenecks that plague various materials towards practical device implementations including quantum light sources. Our review highlights a growing need for excitonic materials development together with optical engineering and imaging techniques to harness the utility of excitons and their host materials for a variety of applications.

## 1. Introduction to Excitonic Materials

Semiconductors are the work-horse materials class of modern microelectronics and opto-electronics. While the electronic devices are mostly reliant on Si, the opto-electronic and photonic devices have been dominated by compound semiconductors. The operating principles of these devices rely on the interaction of light with the crystal medium. In most cases, the interaction is simple propagation of light or absorption of the photon (boson) that results in excitation of a bound electron from valence band into a free electron (fermion) in the conduction band leaving behind a free hole (fermion). However, as the device dimensions shrink, the case of tight integration of opto-electronics and photonics with electronics becomes stronger. From this perspective, it is imperative to understand interaction of light with semiconductors under strong physical confinement. This is a frontier area for basic as well as applied research in nanophotonics and undoubtedly tremendous amount of work has been done and is being done with regards to light confinement in nanowires, nanorods and quantum wells of Si as well as compound semiconductors. However, another approach is to study light matter interactions in semiconductors that are naturally "low-dimensional" or "quantum-confined". The interaction of light with such crystals is fundamentally different since the above gap photon (boson) excites a tightly bound electron-hole pair called the exciton (also a boson). These tightly bound excited state quasiparticles offer a new regime for light confinement in deep subwavelength dimensions. Adopting these materials in devices is expected to drive a transition from opto-electronic devices to opto-excitonic or exciton-photon devices where both light and exciton interact strongly in the same or closely coupled transport media leading to novel device functionalities.

Achieving stable excitons at room-temperatures depends on the dielectric constant of the material that can influence the Coulomb interaction between electron and holes. Further, confinement of excitons leads to an increase in binding energy, which scales with reduction in semiconductor dimensions. In general, 0D semiconductors possess higher exciton binding energy than their 3D counterparts due to confinement. Beyond understanding the functioning of opto-electronic devices, the neutral and quasiparticle nature of excitons have been explored for fundamental studies in solid-state physics and quantum optics. Excitons can be broadly classified into Frenkel and Wannier-Mott excitons based on the electron-hole binding energy as explained in Box 1. Frenkel excitons are observed in organic molecules,[1] molecular assemblies (J-aggregates)[2] and π-conjugated organic chains[3] owing to their low dielectric constant (or strong binding energy ~0.5 to 1 eV) facilitating stable excitons at room temperature. Furthermore, these Frenkel excitons have a small Bohr radius that are localized to the molecule with limited diffusion constant. On the other hand, Wannier-Mott excitons possess a large Bohr radius compared to their lattice constant, high diffusion constant and low exciton binding energy yet stable at room temperature. These excitons are observed in 0D - quantum dots (QDs) namely II-IV QDs, perovskite QDs,[4–6] 1D-single-walled carbon nanotubes (SWCNT),[7–9] and 2D - transition metal dichalcogenides (TMDCs),[10,11] and hybrid organic-inorganic perovskites (HOIP).[12] The excitonic materials covered in this review article are shown schematically in Figure 1. Briefly, QDs are semiconductor nanocrystals with tunable bandgap by varying the nanocrystal size of the same constituents (eg. CdS, $CsPbBr_3$). SWCNTs are carbon allotropes with $sp^2$ hybridized carbon sheets (monolayer graphene ribbons) rolled up into seamless tubes. TMDCs are van der Waals layered materials with a chemical formula of $MX_2$, where M - transition metal (Mo, W) and X - chalcogen (S, Se, Te). HOIPs are lead halide (inorganic - $PbI_2$) quantum wells separated by organic cations (or ligands) stacked to develop a multilayered quantum well material. The number of quantum wells and organic layers in a unit cell defines the order of the HOIP.[12–16] Different from exciton property of individual materials, stacking two semiconducting materials such as TMDC heterostructure,[17–19] TMDC/organic,[20–22] TMDC/perovskite QDs[23–25] and TMDC/HOIP[26–29] hybrid structure or even trilayer heterostructures[30,31] results in the formation of charge-transfer excitons or interlayer excitons which exhibits long-lived states with their dipole moment (out-of-plane) different from their in-plane dipole moment. Recently, careful control over these heterostructures both in and out of plane via twist and stacking degree of freedom have allowed investigation of excitons in Moiré[32–36] and hybrid superlattices.[37,38] Such variety of excitonic behavior present in the multi-dimensional semiconductors as discussed above, can open new avenues of opto-electronic properties when they are coupled with light by integrating them in an external photonic system. For instance, excitonic materials hybridized with light in an optical mirror enable strong light-matter coupling to form exciton-polaritons, which are half-light, half-matter states. These polariton states are pivotal in the realization of Bose-Einstein condensates for polariton lasing,[39] polariton photochemistry for photocatalytic reactions,[40,41] entanglement,[42,43] neuromorphic,[44] and quantum computing[45–47] which are some of the emerging topics for studying exciton physics and realizing novel opto-electronic devices. Without limiting the external medium to a photonic system, these hybrid light-matter states can also be controlled electrically, which expedites their potential for integration into microelectronic systems.[7,48–50]

This review will focus on understanding the light-matter coupling in ensemble of particles (strong coupling) and single-particle (weak coupling) system for both Frenkel and Wannier-Mott excitonic materials of varying dimensionalities ranging from 0D – QDs and organic molecules to 1D – organic chains and nanotubes to 2D – van der Waals materials such as TMDC monolayers and HOIP. We present the role of optical nanocavity mode volume in achieving strong coupling when integrating these excitonic materials and provide a comparison to evaluate their merits in various types of cavities. Second, while several prior attempts at reviewing excitonic materials have been made, their photonic properties have only been discussed in the context of far-field spectroscopy.(51–58) In the past decade, nanostructure engineering in the sub-wavelength limit has become prevalent due to the widespread availability of high-resolution lithography techniques. This has enabled precise engineering of resonant, light-matter interaction at a mesoscopic level by staying agnostic to the electronic structure of the medium and allow tailoring of electromagnetic near-field per will such as by designing various types of nanocavities. Therefore, studying excitonic effects in engineered near-field electromagnetic environments and more importantly by using near-field spectroscopy to spatially map optical signals at deep subwavelength resolution has taken a defining role in shaping this rapidly evolving research area. In the nanocavity mode, the potential of using near-field spectroscopy to understand exciton dynamics, observing forbidden exciton transitions,(59) superradiance in molecular assemblies(60) and studies emerging in far-field technique will also be detailed. Finally, current state-of-the-art device architecture demonstrating electrical control of hybrid excitonic states as well as individual excitonic quantum emitters. will be presented. Figure of merits for materials in both strong and weak coupling regime will be discussed and possible research interest from the present understanding of these excitonic materials will be discussed. Recent developments on extending the application of hybrid state in the artificial lattice,(61–63) optical neuromorphic computing,(44) exciton-polariton transistor(64,65) and photochemistry(40,41) and quantum emitters in Moiré excitons(66) will be outlined.

---

Box 1:

[Terminology in Excitonics]

**Binding energy**: Binding energy of 1s exciton is the difference between electronic and optical bandgap. Nanomaterials usually exhibit high binding energy due to quantum confinement. i.e. In a 2D semiconductor, ~100 meV hence stable at room temperature.

**Decay time**: Time dimension quantity describing how long exciton lasts without recombination. Eventually the exciton decays by either radiative or non-radiative recombination. Radiative decay rate is measured by ultrafast measurement. Typically ~10-100 ps scale and correlated to diffusion length.

**Oscillator Strength**: Dimensionless quantity related to absorption or emission of dipole. Excitonic dipole is simply described by $\Gamma_o/\omega_o$ (where $\Gamma_o$ is decay rate and $\omega_o$ exciton resonant frequency)

**Exciton Radius**: Excitonic Bohr radius which can be described by hydrogen atom model with effective mass of electron and hole. Wannier-Mott exciton and Frenkel exciton.

[Categorized Types of Exciton]

**Trion/Biexciton**: Excited states exciton formed by additional combination with charged carrier or another exciton during photoexcitation.

**Bright exciton/ Dark exciton**: Dark exciton is spin-forbidden optical transition hence it lasts significantly longer than bright exciton (spin-allowed)

**Singlet/Triplet**: In organic materials, singlet (S=0) and triplet (S=1) exciton are observed. Singlet and triplet are analogous to bright and dark exciton, respectively. Triplet lasts $10^4$ times longer than singlet.

---

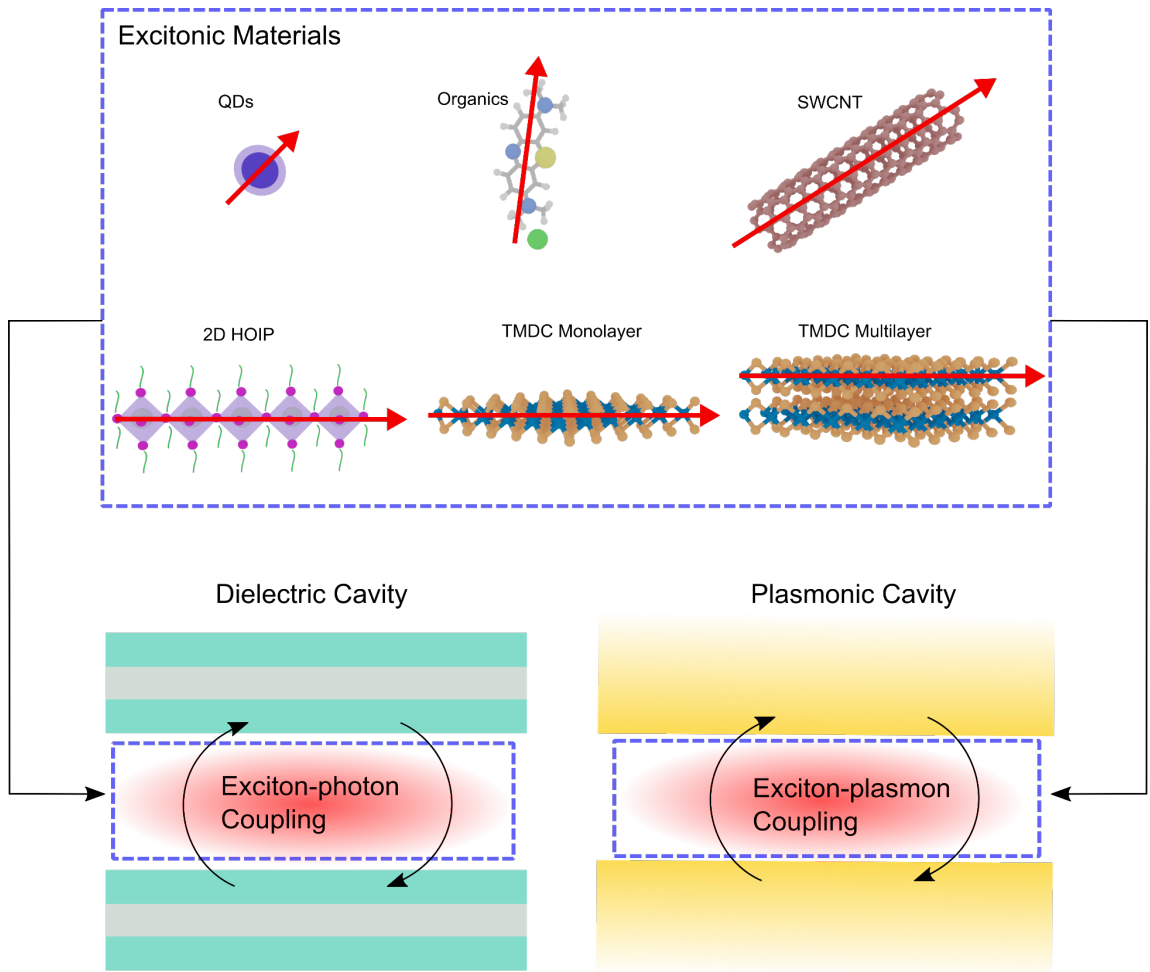

**Figure 1.** Light-matter coupling excitonic materials with an external cavity. Excitonic materials can be from 0D - QDs and organic molecules; 1D - SWCNT; 2D - Hybrid organic-inorganic perovskites (HOIP) and transition metal dichalcogenides (TMDC) monolayer and 3D - TMDC multilayer. The red arrow points to the transition dipole moment ($\mu$) in the excitonic medium. Schematic showing dielectric and plasmonic cavity as an external cavity medium for different excitonic materials placed inside them.

## 2. Strong coupling of excitonic materials in a cavity

Optical cavities ranging from Fabry-Pérot etalons to photonic crystals and plasmonic cavities typically mediate the coupling of photons with excitons into hybrid modes. Ideally, optical cavities are expected to confine light within the cavity for an infinitely long period of time. However, this is far from the reality due to the non-ideal properties of materials and imperfections in cavities. One of the important characteristics of the optical cavity is the quality factor ($Q$), which is a measure of energy decay within a cavity by absorption or scattering to the environment, and the decay rate is denoted as $\Gamma_{cavity}$. Furthermore, the emitter decay ($\Gamma_{emitter}$) also influences the strong coupling in an optical cavity. Therefore, the coupling strength ($g$) must exceed the sum of both cavity decay rate and emitter decay rate ($g > \Gamma_{cavity} + \Gamma_{emitter}$). Taking this condition into account, the quantum emitters (InAs QDs(67) in photonic crystal cavity) are studied at cryogenic temperatures with large mode volume ($V_m$) and high $Q$ factor to achieve strong coupling, as shown in Figure 2a. Although the large microcavity mode volume satisfies the condition for coupling strength to observe strong coupling in these systems, it can be noted that the operating temperature is limited to the cryogenic levels. To achieve room-temperature operation, a high $Q$ factor cavity with reduced mode volume needs to be addressed as it is intrinsically limited by the mode volume ($V_m = \lambda/2n)^3$, where $\lambda$ is the wavelength of light in vacuum and n is the index of refraction of the medium. This reduced mode volume for dielectric cavities remains a challenge even today, with the advanced solid-state device fabrication largely limited by the refractive indices of the medium. On the emitter contribution, the emitter decay rate is also limited by $k_BT$, which requires low Q factor cavities (less than 100). As the condition for strong coupling scales with $Q/\sqrt{V_m}$, this requirement can be met by drastically reducing the mode volume to less than $10^{-5}$ $V_\lambda$ as shown in Figure 2a.

Plasmonic cavities are emerging nanocavity systems that can form small mode volumes and achieve strong coupling despite their low $Q$ factor. One such technique is nanoparticle-on-mirror (NPoM), where the excitonic material is placed on a gold substrate (mirror) and covered with a nanoparticle. When the plasmonic resonance wavelength of the nanoparticle overlaps with the absorption of the excitonic material placed in the nanocavity, the condition for strong coupling is satisfied to form hybrid states. This technique is highly simplified in terms of fabrication while deterministic spatial positioning is still a challenge; however, it is a much more reliable and reproducible system than plasmonic bowtie nanoantennas. Nevertheless, the nanosized nature of excitonic materials permits achieving sub-nanometer mode volumes and strong coupling (Figure 2a). Without limiting to far-field studies, the plasmonic nanocavity can be exploited to study exciton dynamics across the lateral dimension, as discussed in the next section.

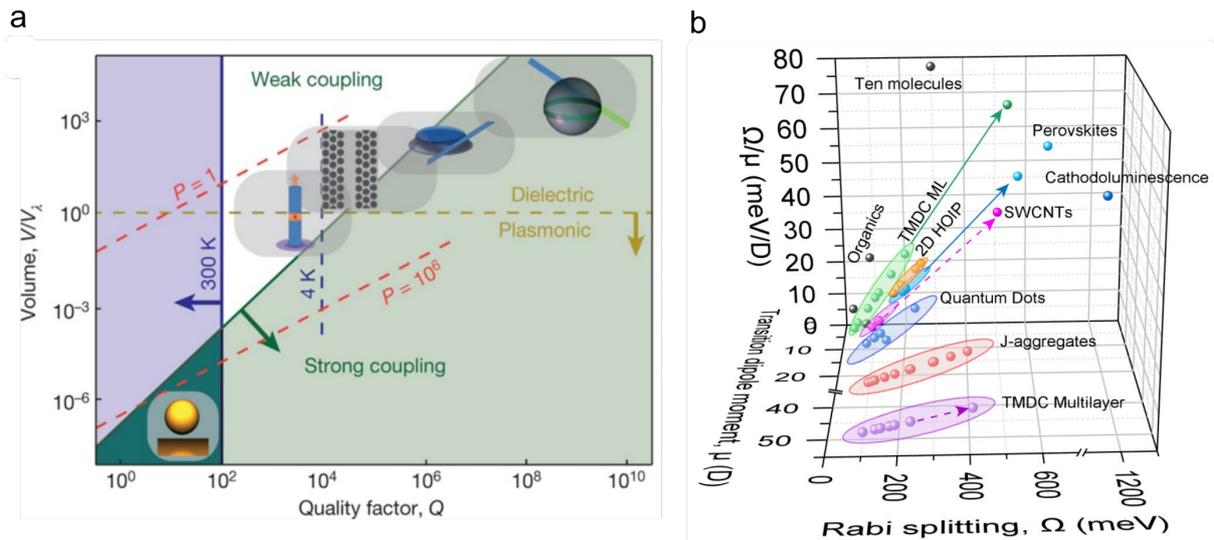

**Figure 2.** Strong coupling of excitonic materials in a cavity. (a) Comparison of different cavity mode (microcavity, resonator, photonic cavity and micropillars) shows that plasmonic cavity can achieve strong coupling at 300 K due to their ultrasmall mode volume despite a low $Q$ factor. (b) Excitonic materials namely QDs,(6,68–72) organic molecules – (0D);(73–75) SWCNT,(76–79) perovskites – 1D,(80–83) J-aggregates,(84–93) 2D HOIP,(94–97) TMDC monolayer (ML)(98–105) – 2D; and TMDC multilayer(99,100,106–109) – 3D studied in cavity mode. Rabi splitting ($\Omega$) to transition dipole moment ($\mu$) as a figure of merit ($\Omega/\mu$) is compared for these excitonic materials (b). Dotted and solid line shows the $\Omega/\mu$ achieved for excitonic materials where gain medium itself acts as an exciton-plasmon cavity and Fabry-Pérot (FP) cavity, respectively. Panel (a) - Adapted from ref.(73) Copyright 2016 Nature Publishing Group.

Excitonic materials from 0D to 1D and up to 3D have different transition dipole moment ($\mu$), which influences the light-matter coupling in a cavity. For reasonable comparison of these materials in a strong coupling regime, the ratio of Rabi splitting ($\Omega$) to transition dipole moment ($\mu$) is considered as a figure of merit in this review and shown in Figure 2b. It should be noted that, although the photonic cavity has a very high $Q$ factor, and the plasmonic cavity has small mode volumes, the outcome of Rabi splitting which is dependent on $Q/\sqrt{V_m}$ will be dominated by plasmonic cavity when using $\Omega/\mu$ as a figure of merit.(53) Nevertheless, in this review, our emphasis in Figure 2b is on the performance metric for different excitonic materials, which is mostly dictated by their transition dipole moment and staying agnostic to cavity type. Second, self-hybridized light-matter coupling without an external cavity is an upcoming field of interest in the community.(107–109) For instance, a self-hybridized system – perovskite nanowire (1D) with an intrinsic Fabry-Perot mode shows a Rabi splitting, $\Omega$ ~ 564 meV in the strong coupling regime.(80) In such a case, although the $\mu$ of the system is less than 20 $D$, achieving strong coupling in a microcavity is a remarkable step. This is due to various factors such as refractive index ($n$), exciton oscillator strength ($f$), and $V_m$ affecting $g$ in the F-P cavity ($g \propto \sqrt{(n \cdot \frac{f}{V_m})}$). Nanowires with sub-wavelength dimensions reduce the mode volume drastically and hence enhance the coupling strength. On the other hand, patterned TMDC multilayer with high oscillator strength also shows a strong coupling ~400 meV.(110) For such systems, there is a large spread in oscillator strengths for different dimensionalities of semiconductor materials. Reduced mode volume is not reported in several published reports of Rabi Splitting likely due to difficulty in estimating accurate field profiles and mode dimensions on such small scales. Here, we will focus on a critical analysis using $\Omega/\mu$ as a figure of merit to identify the best strategies for achieving strong coupling with photons in this review article.

0D materials such as organic molecules and QDs support Frenkel excitons and Wannier-Mott excitons, respectively.(52,111,112) Evidence of strong light-matter coupling has been presented in organic molecules such as methylene blue positioned in plasmonic nanocavities with their dipole moment along the cavity electric field. Likewise, QDs with random dipole moments that are deterministically placed in plasmonic nanocavities such as bowtie antennas have also shown evidence of strong coupling.(6,68–71,113) Single-particle probe techniques such as cathodoluminescence were used to probe especially large Rabi splittings (~1200 meV) from individual CdZnS/CdS QDs placed in Al nanocavities.(71) For a detailed account on the strong light-matter coupling in inorganic QDs measured in the far-field, the readers are referred to prior published review articles.(52,53)

Akin to inorganic QDs, functionalized SWCNTs have also shown evidence of strong light-matter coupling from the reflectance spectroscopy studies.(76,77,79) SWCNTs placed in plasmonic cavities or etching SWCNT films/bundles into nanoresonators have shown Rabi splitting ~100 meV(76,77) and ~500 meV,(79) respectively. Also, polariton emission from the hybrid states in SWCNT was observed in the telecommunication wavelength (1330 nm), which is tunable by varying the SWCNT chirality. Further, the high charge carrier mobility and cavity-free polariton emission render SWCNTs promising for electrically injected polariton lasers.(7) However, SWCNTs have been plagued with poor PL emission efficiencies in a thin-film form. Overcoming this hurdle is of fundamental importance and presents a research challenge in materials optimization.(79,111,114)

In contrast with SWCNTs, J-aggregates are visible-NIR emitting self-assembled molecular assemblies where the dipole moment of an individual molecule is coupled, which leads to extended delocalization of excitons throughout the aggregate length (theoretically).(2,115–117) Such strong dipole-dipole coupling in J-aggregate results in increased net transition dipole moment and oscillator strength with narrow absorption/emission linewidth. Strong light-matter coupling has been observed by optically pumping J-aggregates in DBR cavity,(118) metallic dimers/gratings,(119–121) gold gratings,(92) Si nanoparticles as Mie scatters,(122) dye coated plasmonic particles,(120,123–128) and particle on the metallic substrate(75,129) as well as electrical injection in J-aggregate/GaN microcavity up to 200 K.(130) In these studies, the J-aggregates with excitonic absorption peak between 2.25 - 2.48 eV form hybrid states with silver plasmonic particles.(120,123–128) Although silver supports visible plasmon, the narrow interband transition in aluminum (~1.5 eV) allows it to extend the surface plasmon response up to the ultraviolet range and also as a low-cost alternative to silver.(131)  Consequently, replacing silver plasmonic particles with aluminum pads and covering them with J-aggregates doubled the Rabi splitting (400 meV). In addition to hybrid state formation, the strong coupling of molecules in a cavity has inverted single/triplet state in organic molecules,(132) delayed fluorescence,(133) and suppressed photo-oxidation.(134) These reports point toward evidence that strong coupling of light and excitons can alter the exciton recombination pathways, which in turn influences the opto-electronic device performances as well as photochemistry.

In contrast with 0D and 1D semiconductors discussed above, 2D excitonic TMDCs offer gate (electrostatic) tunability of exciton-trion populations, thereby suppressing trion formation (non-radiative decay channels) and enhance the quantum yield of the excitonic emission.(135,136) Such gate-tunability also controls the formation of exciton-polaritons vs. trion-polaritons for TMDCs placed in a plasmonic cavity.(101) Besides, TMDC monolayer with broken symmetry introduces valley polarization, which is enhanced upon integrating TMDCs with chiral photonic metastructure arrays.(137) Such property of selective excitation of band valleys in TMDCs by circularly polarized light adds another knob for tuning light-matter interactions when studied in the strong-coupling regime.(137,138) Finally, the atomic flatness and nanoscale thickness of TMDCs renders them particularly suitable for integration into NPoM architecture for strong coupling studies.(99,100) In fact, the $\mu$ of TMDC monolayer (5 D) is lower by three-orders of magnitude compared to the plasmonic Au nanoparticles (1.3 x $10^4$ D). This stark difference motivated the researchers to combine Au nanoparticles with $WS_2$ monolayer inside the plasmonic cavity. The collective coupling between plasmon-exciton mode in resonance with the cavity mode led to enhancement of the Rabi splitting from 75 meV (without plasmonic particles) to 535 meV (with plasmonic particles).(105) Nevertheless, such a hierarchical design will have a small excitonic component in the hybrid state despite reaching ultrastrong coupling. This study can potentially serve as a route to circumvent the intrinsically low oscillator strengths (for example, monolayer TMDCs or organic chromophores) in excitonic materials that have hindered strong coupling. HOIPs are another class of emerging 2D excitonic media – have also shown strong coupling upon integration in DBR cavity(95) or in metasurfaces on $SiO_2$.(139) Thick HOIP single crystals (without any external cavity) have also shown similar Rabi splitting (170 meV). Cavity-free light-matter hybridization in excitonic media have been discussed in detail later.

From the above discussion, we have summarized the strong coupling of excitonic materials from far-field studies. We have discussed several aspects of the strong coupling from exploring plasmonic cavities, intrinsic cavity mode to probing hybrid light-matter states in various optical cavities. In the next section, we will discuss about near-field spectroscopy techniques in which we attempt to highlight new physics that is obscured in far-field spectroscopy.

## 3. Near-field spectroscopy as a probe for exciton-photon coupling

Near-field spectroscopy involves an NPoM or metal probe to collect local optical information with a resolution < 20 nm and ultralow cavity mode volumes containing the excitonic medium. Most importantly, scanning near-field techniques provide direct evidence of new photophysics both with high spectral and deep sub-wavelength spatial resolution in Raman(140,141) and photoluminescence (PL) spectrum.(142) Near-field spectroscopy in plasmonic nanocavities for organic molecules were investigated for methylene blue (0D)(73) and 1D zinc phthalocyanine (1D)(60,143) are reported. Encapsulating methylene blue in a Cucurbit[7]uril matrix and placing a plasmonic Au particle on top, strong coupling was achieved (Figure 3a).(73) In this study, the relation between coupling strength ($g$) and number of molecules ($N$) in the cavity following: $g \propto \sqrt{N}$ was demonstrated. The orientation of dipole moment of the methylene blue molecule along the electric field of the cavity led to enhance the coupling strength up to 200 meV with increasing the number of molecules in the nanocavity (Figure 3b).

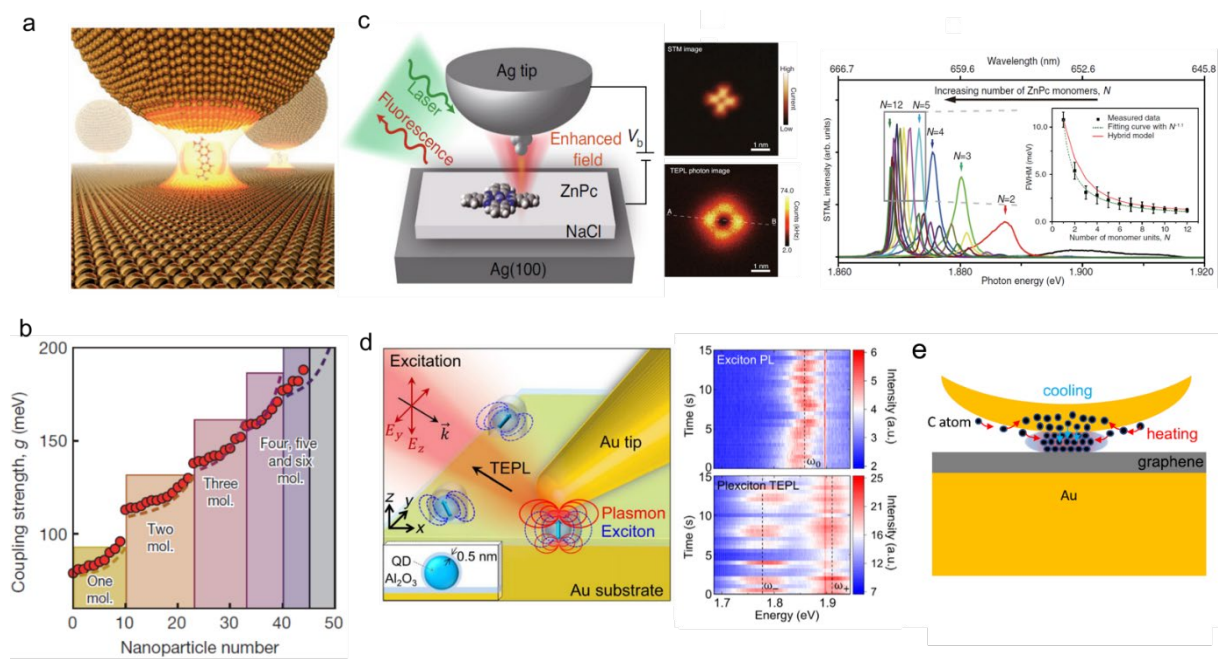

**Figure 3.** Near-field spectroscopy for 0D and 1D excitonic materials in strong coupling. (a,b) Ultrasmall mode volume in a plasmonic cavity for strong coupling organic molecules showing an increase in coupling with increase in number of molecules. (c) Scanning tunneling microscopy induced luminescence (STML) for zinc phthalocyanine (ZnPc) molecular chain in a plasmonic nanocavity showing imaging at molecular level. Appearance of superradiance peak and linewidth narrowing (inset) for different ZnPc chains (right panel in c). (d) Near-field spectroscopy set-up to study QDs in strong coupling regime (left panel in d) and photoluminescence spectrum of exciton under weak coupling and plexciton under strong coupling (right panel in d). (e) Carbon quantum dot formation and plexciton states under nanocavity mode. Panel (a,b) - Adapted from ref.(73) Copyright 2016 Nature Publishing Group. Panel (c) - Adapted from ref.(143) Copyright 2020 Nature Publishing Group and ref.(60) Copyright 2019 American Physical Society. Panel (d) - Adapted from Ref.(69) Copyright 2019 American Association for the Advancement of Science (AAAS). Panel (e) - Reprinted with permission from Ref.(144) Copyright 2020 American Chemical Society.

Beyond light-matter coupling and achieving hybrid states, the near-field spectroscopy can also help unravel the fundamentals of dipole moment characteristics, its coupling with the electric field and orientations. Generally, the transition dipole moment of the molecules is assumed to be the point dipole moment (PDM) in nature. Using tip-enhanced photoluminescence (TEPL), the ZnPc molecules can be imaged up to sub-nanometer resolution (~8 Å) with a silver tip.(143,145,146) Enhanced PL from near-field in a plasmonic cavity was used to map ZnPc PL with high sensitivity and resolution. As shown in Figure 3c (right panel), the TEPL signal from ZnPc shows four lobes signifying four-fold symmetry of the molecule.(143) Direct observation of the two orthogonal dipole moments that give rise to four-fold symmetry and not PDM is fundamentally novel insight revealed by near-field spectroscopy. The dark region in the center was due to the cancellation of the dipole symmetry from the whole molecule. Further, closely packing these ZnPc molecules in 1D nanowire leads to dipole-dipole coupling. Again using TEPL, red-shifted PL peaks with reduced linewidth as well as signatures of superradiance was observed upon increasing the number of molecules.(60)

Although near-field spectroscopy using TEPL and far-field surface enhanced Raman spectroscopy (SERS) are extensively investigated, unexplored domains such as tip-enhanced Raman spectroscopy (TERS) can be used to identify the non-radiative decay channels, which are dark state for far-field spectroscopy. For instance, using TEPL for nanoimaging and identifying non-radiative spots can also be misinterpreted from photobleaching of organic molecules in ambient conditions. However, the Raman spectrum can show distinct signature of individual molecules (monomers) and J-aggregates.(147–149) In this way, monomer molecules which act as exciton trap states can be identified to understand the spatial position of the molecules in large size assemblies using TERS.(150) Further, introducing time-resolved measurements in near-field technique can be helpful to identify the exciton localization and non-radiative recombination at these exciton traps.

QDs are one of the best model systems to investigate strong and weak coupling due to high quantum yield and its random dipole orientation. Since gap mode creates a high electric field along z-direction, a QD with excitonic dipole along z-direction can strongly couple with plasmonic nanocavity while the QD with other orientation can only induce weak coupling. For example, exciton-plasmon polariton coupled state, so called plexciton, was reported using CdSe/ZnS QD on Au with 0.5 nm of dielectric capping layer (Figure 2d).(69) Multiple QDs showed different types of spectra with respect to degree of coupling between QD excitonic dipoles and gap plasmon modes. In the strong coupling regime, the coupling strength g was found to vary from 70 meV to 163 meV, which reveals plexciton formation due to randomly distributed dipole orientation of each QDs. In the case of weak coupling, on the other hand, only photoluminescence enhancement was observed due to the increased radiative rate by the Purcell effect. Similarly, an in-plane plasmonic resonator tip and was also investigated with CdSeTe/ZnS QDs embedded PMMA medium to observe string coupling wherein a plasmonic substrate is not necessary.(6) Likewise plexcitonic states have also been reported in carbon quantum dots (CQDs) by placing graphene between Au nanoparticle and Au substrate (Figure 2e). In this case the graphene inside the nanocavity acts as a template for forming CQDs resembling the graphene lattice. High electric field in the gap mode triggers carbonization of the pre-adsorbed hydrocarbons on the Au tip into CQDs. The authors reported anti-crossing (Rabi splitting) of ~200 meV in emission revealing successful formation of plexcitonic states.(144)

## 4. Near-field spectroscopy study on 2D materials

Near field spectroscopy is particularly helpful for investigating 2D semiconductors since it is atomically thin in thickness and can be infinitely large in lateral size. This 2D structure allows it to maintain uniformly confined plasmonic gap modes through the

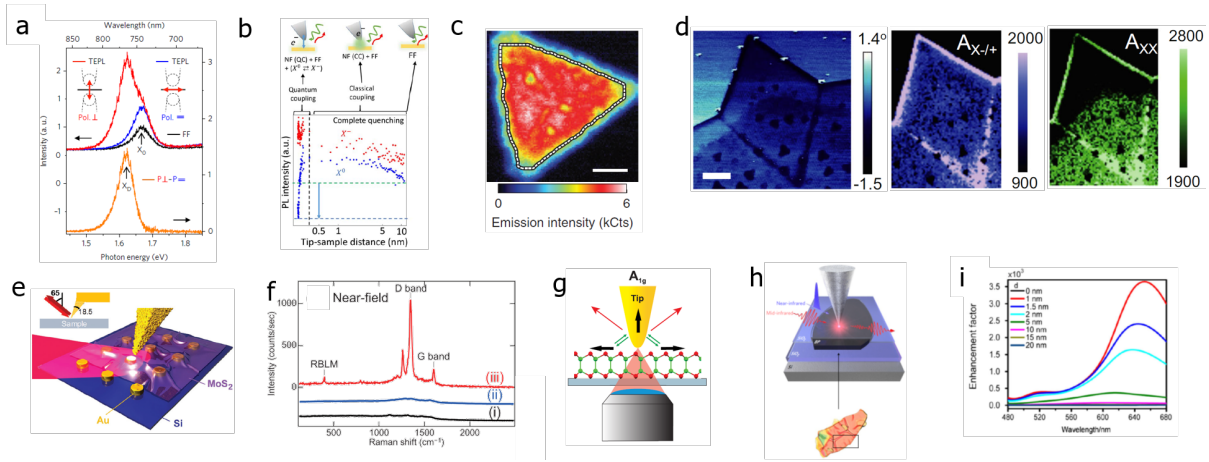

**Figure 4.** Near-field spectroscopy for 2D materials. (a) Radiative decay of dark exciton on WSe$_2$ by Purcell effect. (b) Trion/neutral exciton control on WS$_2$ using plasmonic picocavity. (c) Nanoscale spatial variation of photoluminescence on MoS$_2$, exhibiting distinct interior and edge emission. (d) Correlation between KPFM phase, trion and biexcitonic emission mapping of MoSe$_2$. (e) TERS spectrum of MoS$_2$ on Au nano cylinder array. (f) TERS spectrum of graphene nanoribbon (red), Polyanthrylene/Au (111) (blue) and clean Au (111) (black), respectively. (g) TERS spectrum of GaS to investigate phonon correlation length. (h) Ultrafast switching of interface polariton on 2D black phosphorus. (i) FDTD simulation on TEPL of black phosphorus under nanocavity. Panel (a) - Adapted from Ref.(69) Copyright 2019 American Association for the Advancement of Science (AAAS). Panel (b) - Adapted from Ref.(151) Copyright 2019 American Association for the Advancement of Science (AAAS). Panel (c) - Adapted from Ref.(152) Copyright 2015 Nature Publishing Group. Panel (d) - Adapted from Ref.(153) Copyright 2020 Nature Publishing Group. Panel (e) - Reproduced with permission from Ref.(154) Copyright 2018 The Royal Society of Chemistry. Panel (f) - Adapted with permission from Ref.(155) Copyright 2014 American Chemical Society. Panel (g) - Reprinted with permission from Ref.(156) Copyright 2019 American Chemical Society. Panel (h) - Adapted from Ref.(157) Copyright 2017 Nature Publishing Group. Panel (i) - Reprinted from Ref.(158) Copyright 2019 Wiley Publishing group.

entire flake. Within 2D materials, excitonic TMDCs have intrigued highest interest in the near-field spectroscopy studies due to their distinct optical properties and emergence of locking in spin and valley degrees of freedom,(159,160) presence of spatial heterogeneity,(154,161,162) strain(163) and in-plane epitaxial hetero-layer structures.(164–167)

In semiconducting TMDCs, strong spin-orbit coupling leads to spin and energy separation of the conduction and valence bands. For example, Mo based TMDC forms the lowest conduction band with spin-up states while W based TMDC does with spin-down states. Since the highest valence band consists of spin-up states, the smallest energy optical transition is spin-forbidden for W based TMDC, which is termed as dark exciton. Dark excitons have a long decay time is nanoseconds since they do not couple easily to emit light and hence are the subject of numerous investigations for quantum information processing.(168) In contrast  the bright excitons in TMDCs have a lifetime of picoseconds.(169,170) However, using a plasmonic tip in near-field spectroscopy, allows one to turn on radiative decay and control of dark and bright excitons in 2D TMDC crystals. Specifically, the plasmonic tip enables to directly prove the existence and radiative control of dark exciton emissions in $WSe_2$ crystals due to its out-of-plane optical transition (Figure 4a).(171) The radiative decay of dark exciton requires a spin-flipping process typically induced by a large magnetic field and records $10^2$-$10^3$ times smaller intensity as compared to the bright exciton.(172) In vicinity of the tip, however, the dark exciton is more emissive due to the increase in excitation rate of dark exciton by out-of-plane field enhancement as well as enhanced radiative decay rate due to the Purcell effect. Further it controls dark exciton emission by changing the plasmonic gap between 1 to 5 nm. As the tip-sample distance decreases, high emission was achieved due to the enhanced quality factor of the gap mode. Further, by controlling the cavity in the picoscale regime, it was able to control charged exciton (trion) formation (Figure 4b). Under this small separation between the tip and substrate that forms a picocavity for plasmon confinement, the tip can also tunnel charge into the $WS_2$, which leads to trion formation.(151) These results verify that a scanning plasmonic tip can form an active cavity with small mode volumes that can allow control over various types of excitonic emissions from 2D TMDCs.

Nanoscale spatial resolution in near-field spectroscopy is further important in resolving crystal defects and interface heterogeneities in 2D TMDCs. The technique has been used to probe features such as defects, grain boundaries[161,173] disordered edges,[173] oxidized surface by aging effects,[174] electronic contact quality[175] and strain.[154,176–178] Although these structural heterogeneities are vital for understanding 2D semiconductors and optimizing opto-electronic devices comprising them, they cannot be resolved by diffraction-limited far-field spectroscopy tools due to their physical sizes and interaction with visible frequency photons. Hence near-field spectroscopy plays an essential role in investigating these nanoscale features. The earliest reports of near-field spectroscopy use aperture type campanile tips which can resolve by ~ 60nm features revealing differences between disordered edge emitting regions and interior regions in CVD grown monolayer $MoS_2$ (Figure 4c).[173] Specifically, ~300 nm of the peripheral edge region was discovered with disordered, low intensity, broad energy emission while the interior region showed higher energy and intensity emission due to reduced trion formation. Inspired by this initial study, similar studies have been conducted with apertureless plasmonic tip. Since the apertureless tips utilize the Purcell effect to confine high electric fields in nanocavity, tip-radii determines the resolution which is typically ~20 nm. Using the apertureless tip, similar spatial variation in photoluminescence in CVD grown monolayer $MoS_2$ crystals was attributed to point defects, nanoscale terraces, and edge emission.[179] For example, localized exciton emission from nanoscale bubbles of monolayer $WSe_2$ with diameters of ~50 nm was observed.[178] Because bubbles induce a slightly smaller optical bandgap surrounded by continuous excitonic medium of TMDCs, exciton funneling effect arises, leading to brighter emission at the bubble. Strain, which is an important parameter of the bubbles was quantitatively estimated by solving strain tensors using topography and Poisson's ratio, which coincides with strain evaluation from TERS.(178) Due to the operation at room temperatures, the strained nanobubbles in 2D TMDCs are particularly interesting from the perspective of single photon emission as discussed later. Another example of 2D excitonic semiconductors where near-field spectroscopy is particularly useful is on artificially grown lateral heterojunctions and superlattices of TMDCs like MoSe$_2$/WSe$_2$.(165,180,181) Owing to similarity in lattice constants and crystal structures, combined with their atomically-thin structure it is feasible to make in-plane stitched, lateral hetero-interfaces between two different 2D semiconductors of varying band gaps. The junction comprises of depletion region as well as strain which are of deep sub-wavelength dimensions. In addition, the as-grown lateral MoSe$_2$/WSe$_2$ heterostructures exhibit sub-wavelength periodic bandgap changes of MoSe$_2$ and WSe$_2$. Using TEPL profile, the junction width was estimated as ~150 nm which is sub-wavelength scale.(165,174,180) TEPL analysis has also provided evidence of directional hot-electron injection from WSe$_2$ to MoSe$_2$ by potential gradient at the junction, which results in enhancement of luminescence in MoSe$_2$ region. Another big advantage of using plasmonic metal tip is a correlation with non-destructive electrical measurements such as potential, capacitance and electric- and photo- conductive measurement. Correlation between TEPL, TERS, potential and second derivative capacitance maps on MoSe$_2$ revealed nanoscale inhomogeneities.(161) The capacitance change under illumination/dark environment also proved that grain boundaries can be metallic twin boundaries.(182) Further, optical properties of oxidized MoSe$_2$ monolayer have also been verified by correlating KPFM phase to photoluminescence map (Figure 4d).(153) Definitive proof of oxidation induced dielectric environment change and doping which results in enhanced biexciton formation was obtained from this correlation between optical and electronic scanning probe measurements. Scanning probe techniques coupled with light illumination have also paved the way for probing buried interfaces of excitonic materials such as TMDC-metal contacts where the TMDCs are directly exfoliated on flat metals. When combined with local photocurrent measurements and resonant Raman spectroscopy recent studies have shown spatial inhomogeneity in the $WSe_2$-metal contact interface.(183) The probing of TMDC-Metal buried contact interfaces can be further extended to include even directly evaporated contacts via a unique Au-assisted transfer technique.(184) This transfer technique flips the buried TMDC-metal interface, exposing it to the sample surface. Further, the contact interfaces created by direct metal evaporation are clearly distinguished from typical TMDC-metal interface fabricated by direct exfoliation process via scanning conductance, potential, photoluminescence and Raman probes. This has allowed comparison of TMDC-metals contacts that are not only formed via different evaporation/exfoliation schemes but also allowed investigation of different TMDC/metal combinations in determining the lowest resistance contact making schemes.(175)

Near-field Raman spectroscopy is also an important spectroscopy mode for studying 2D TMDCs since it provides complementary information as compared to PL such as nanoscale inhomogeneities in charge doping, stoichiometry, and strain. For instance, TERS measured from polycrystalline $MoSe_2$ transferred on Au substrate showed nanoscale variance in $A'_1$ and $E'$ mode intensity at basal plane and grain boundaries.(161) This variance was attributed to local doping change as a result of oxidation of the selenide. Since high signal to noise ratio is challenging to obtain in TERS, various types of surface preparation techniques have been developed to enhance TERS signals in practice. For instance, certain TERS studies of $MoS_2$ on Au nano arrays have used as cylindrical (Figure 4e) or triangular structures/shapes of Au.(154,176) An 8-fold enhancement of TERS signal at the edge of the Au clusters was observed. Under the non-resonant TERS conditions, a large intensity enhancement of the out-of-plane $A_{1g}$ mode was found while the in-plane $E_{2g}$ mode was suppressed due to the out-of-plane E-field enhancement in the nanocavity.

Graphene is the prototypical 2D material which exhibits semi-metallic properties with no bandgap. However, confining graphene to < 10 nm dimensions opens up a bandgap and enables it to support excitons, akin to carbon nanotubes.(185) Accordingly, near-field optical imaging has not only revealed defects and doping in graphene,(186,187) but also properties in confined graphene such as edge induced localized electronic states.(188) Graphene nanoribbons (GNRs) have been extensively studied due to their size-dependent bandgaps.(189–191) However, due to their small sizes, the edge states dominate transport in GNRs which motivated the investigation to study edge states of the nanoribbons via near field spectroscopy (figure 4f).(155) The edge states were investigated by spatially mapping the D peak intensity, which is mainly arises from the armchair type edges.(192) The phase-breaking length, which represents the average distance of excited electron-hole pair diffusion, was observed to be 4.2 nm which was a far smaller value as compared to the estimate from far-field measurement (~50 nm). This discrepancy is once again a result of the large beam size of the Gaussian beam of the far-field measurement, which is not able to resolve ~4 nm width and therefore highlights the power and utility of near-field, scanning optical nanoprobes.

Aside from TMDCs of Mo, W and GNRs, group III-VI layered semiconductors MX (M = Ga and In; X = S, Se and Te) are also important in the layered materials family, showing relatively large optical bandgaps for monolayers (~2.5-3.0 eV).(156,193) These large optical bandgaps have attracted their use in new applications such as blue LEDs.(194) However, low emission efficiencies and weak Raman signals in monolayers presents challenges in investigating them via far-field spectroscopic techniques.(194,195) Therefore, tip-enhanced near-field spectroscopy or coupling with plasmonic arrays can offer a viable strategy towards enhancing Raman or PL signals from these materials.(156,196) In this regard, a recent study reported estimating phonon correlation lengths ($L_c$) of 50- 60 nm in few-layered GaS with plasmonic Au tip (Figure 4g).(156) $L_c$ was measured by collecting the Raman spectrum as a function of tip-sample distance. Although two $A_{1g}$ modes have identical symmetry, high order phonon-phonon coupling in $A_{1g}^2$ leads to difference in less TERS enhancement. This result is striking difference from graphene since phonon-phonon coupling as well as mode symmetry influence the Raman scattering in GaS. Near-field studies on III-VI layered semiconductors is currently in early stages. Further, the exciton binding energies in this class of 2D semiconductors is small.(197) Hence it is difficult to observe strong exciton-photon coupling phenomena at room temperature in near-field conditions. Nonetheless they are valuable from an application perspective since they maintain direct bandgap even in the few layers to bulk thickness limit. Therefore, nano-resolution TEPL, defect, strain analysis will be helpful in understanding their optical and electrical properties and develop future applications out of them.

While III-VI layered semiconductors are on the wider gap side, 2D black phosphorus (BP) is another promising layered van der Waals semiconductor for opto-electronics, albeit on the narrow gap side. Its bandgap is tunable from 0.3 eV in the bulk to 1.5 eV for monolayers which covers visible to infrared ranges,(198,199) and has been shown to have applications in terahertz photodetectors(200) as well as shown to have a plasmonic response.(201) Interface polaritons in multilayer BP with ultrafast photo-switching of ~50 fs was reported by using scattering-type near-field with infrared laser (Figure 4h).(157) While experimental studies in near-field spectroscopy in BP are not available yet, a recent calculation predicted enhancement factors up to $10^8$ and $10^3$ for Raman and PL respectively in the BP by using 3D- finite-difference time-domain (FDTD) method (Figure 4i).(158) Still, investigation on BP is still in the early stages. Monolayers present a unique opportunity to study exciton-photon coupling in the near-field particularly given the in-plane anisotropy of the excitons(202) but the fast oxidation rates under ambient conditions require encapsulation which presents significant experimental challenges.

In the above section we have reviewed and discussed how near-field spectroscopy studies, reveal new insights from multi-dimensional semiconductor materials particularly with regards to exciton-photon interactions and coupling. New insights into exciton dynamics at the nanoscale level unraveled by the high spatial resolution of TEPL and TERS highlight the importance of near-field studies.

## 5. Far-field spectroscopy in open cavity systems

While the tip serves as a cavity or resonator for all the near-field investigations discussed above, an external cavity is not necessarily required to exploit light-matter coupling. Recent advances have shown that the gain medium can itself act as an open cavity and negates the need for an external cavity. Extraordinary optical

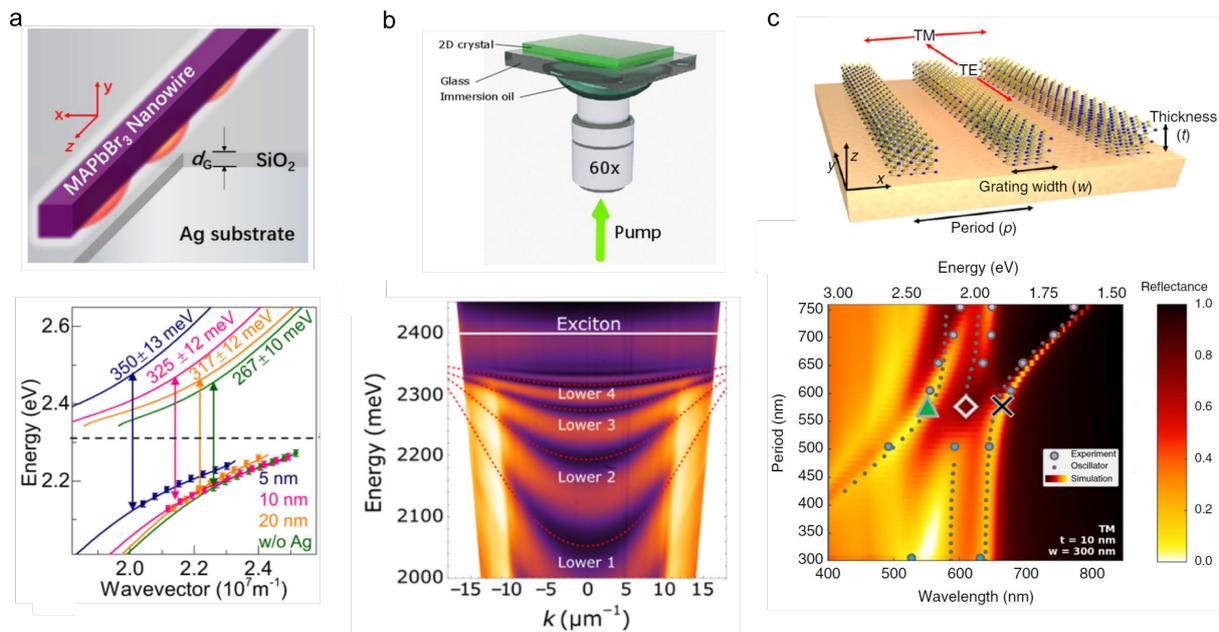

**Figure 5.** Far-field spectroscopy studies for excitonic materials in open cavity systems. (a) Perovskite nanowire in an open plasmonic cavity. Rabi splitting was observed from strong coupling by decreasing the distance between the perovskite nanowire and silver cavity. (b) 2D HOIP single crystal on glass substrate showing Rabi splitting. (c) Multilayer WS$_2$ grating structure (thickness, t = 10 nm, width, w = 300 nm) with varying the period (p) on gold substrate showing strong coupling with increasing WS$_2$ thickness. Panel (a) - Reprinted with permission from ref.(80) Copyright 2018 American Chemical Society. Panel (b) - Adapted from ref.(95) Copyright 2019 American Association for the Advancement of Science (AAAS). Panel (c) - Reprinted with permission from ref.(107) Copyright 2020 Springer Nature CC BY.

constants inherent in materials such as perovskites and TMDCs can sustain F-P (open) cavity mode, which leads to strong light-matter coupling without any external cavity in sub-wavelength thicknesses of these media. Perovskite nanowires on an Ag substrate have shown strong Rabi splitting due to a combination of reduced mode volume and high oscillator strength (Figure 5a, bottom panel). Without limiting them to 1D nanowires, HOIP in 2D single crystals also shows Rabi spitting in an open cavity (Figure 5b). Thick HOIP crystals can sustain polariton and interact strongly with exciton (3 ± 0.5 μeV μm$^2$) at room temperature. Further, the presence of Rashba effect in HOIP introduces spin-dependent polaritons which offer interesting new avenues in spintronics and electrically controlled polariton formation devices.(203–206)

Multilayer TMDCs is another exciting material to realize cavity free strong coupling of excitons and photons. Several recent studies have shown that metastructures such as disks,(108) gratings(107) or bottom up synthesized nanotubes(207) can help support coupled exciton-photon states. These metastructures form an F-P cavity mode and interact with excitons to create hybrid states.(107,108,207) Shaping thick $WS_2$ layers into disks forms anapole modes while free-standing thick films support F-P modes that couple to excitons.(108,208) When placed on reflecting metals such as Au, Ag or Al the thickness of these multilayer $WS_2$ flakes that couple to the F-P (cavity) mode can be substantially reduced (< 20 nm) which concurrently induces near unity absorption in these ultrathin semiconductor samples. Further, etching gratings in these multilayer $WS_2$ crystals leads to excitation of a dielectric-grating mode in addition to a surface plasmon guided mode. Excitons strongly interact with cavity mode and splits into upper exciton-polariton (UEP) and lower exciton-polariton (LEP). Subsequently, the dielectric grating mode interacts with UEP and LEP to form three-oscillator couplings with Rabi splitting ~410 meV.(107) By increasing the grating period, the hybrid modes are tuned, where the grating mode emerges into the non-dispersive upper polariton branch (UPB), while LPB switches from non-dispersive to plasmonic mode. (Figure 5c, bottom panel). Although the multilayer TMDC flakes can show strong Rabi splitting, the indirect bandgap nature limits them from polariton emission, which needs further optical engineering in this direction.(209)

In summary, open cavity and intrinsic cavity systems are very appealing since they allow design and realization of monolithic exciton-photon optical elements.

Further, their design simplicity also opens door to electrodes integration for electrostatic tuning or electrically injected emission. Another important attribute of the open-cavity systems is their accessibility to other media. For example, for driving polariton induced chemistry or catalyzing a reaction, an open cavity system allows easy access for the reactants and has the potential for scale up to larger volumes and length scales for photochemistry and photocatalysis. This is potentially difficult to achieve in a closed, nano-cavity system. Given that several excitonic materials such as SWCNTs, perovskites and multilayer TMDC have shown feasibility in supporting polariton modes in open cavities, this area of research is promising. However, there are several challenges with the concerned excitonic materials such as, their intrinsic low quantum yield, stability under ambient conditions etc. This will limit their applicability to very specific class of devices or other applications. Further, material quality and uniformity as a whole needs to improve and thus there is plenty of more work to be done in that regard. Finally, understanding of electrical contacts to these materials and developing strategies for stable and reliable contact making will be important moving forward as it will facilitate seamless integration into active devices.

## 6. Electrical control of emission from hybrid states

Hybrid states formed upon optical pumping of excitonic materials in an external cavity can be generated externally by carrier injection in the semiconductors. One of such architectures is light emitting field effect transistor (LEFET). Possibility of long-range charge transport in the semiconductors and emission in the center of the channel makes LEFET more efficient emitters than diode architecture where excitons are quenched at the metal electrode. LEFET was fabricated by dispersing

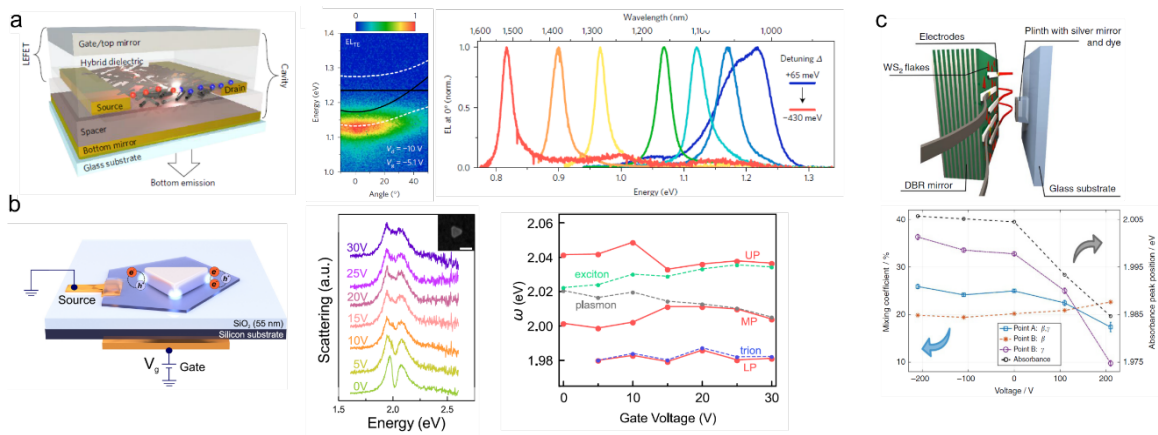

**Figure 6.** Electrical control of strong light-matter coupling. (a) Device structure of bottom contact/top gate light emitting field effect transistor (LEFET) using (6,5) SWCNTs. Angular reflectivity of the channel area in the LEFET showing Rabi splitting in electroluminescence (middle panel in a) and angle-resolved electroluminescence with detuning the cavity (right panel in a). (b) Hybrid plasmon-exciton in monolayer $WS_2$ under electrostatic gating. Dark field scattering showing hybrid state tunability with gate voltage (middle panel in b). Eigen energies of the coupled plasmon-exciton-trion Hamiltonian vs gate voltage at 300 K and atmospheric pressure. (c) Electrically tuned hybrid $WS_2$-organic system in a dielectric microcavity showing polariton composition (fraction of Frenkel and Wannier Mott excitons) along with absorbance at different voltage. Panel (a) - Adapted from ref.(7) Copyright 2017 Nature Publishing Group. Panel (b) - Reprinted with permission from Ref.(210) Copyright 2020 American Chemical Society. Panel (c) - Adapted from Ref.(49) 2017, Nature Publishing Group CC BY.

(6,5) SWCNT in a polymer matrix and sandwiching between DBR mirrors with the bottom electrode and top gate (Figure 6a). Polariton emission emerging from LEFET with Rabi splitting ~125 meV can be tuned from 1,060 nm to 1,530 nm by detuning the cavity (middle panel, Figure 6a). While polariton emission was evident, polariton lasing was not observed even at high injection currents.(7,211)

Electrostatic gating is a widely used parameter for controlling device performance in microelectronics. High photoluminescence quantum yield was achieved for TMDC monolayer in cavity free mode by varying the exciton and trion proportion using electrostatic gating.(136) In a similar architecture, adding a plasmonic nanoparticle on the TMDC monolayer ($WS_2$) can introduce exciton-plasmon hybrid states (Figure 6b).(101,210,212) Tuning these hybrid state with electrostatic gating at different gate bias was demonstrated from reflectance spectroscopy. At 77 K, plasmon-exciton becomes dominant (V= -12 V), however, exciton remains dominant at room temperature (V = 30 V) (Figure 6b). On the other hand, multilayer TMDC with plasmonic particles shows an upper polariton and dispersionless collective mode with a tunable polariton bandgap.(48)

Forming a hybrid exciton state by mixing Frenkel and Wannier-Mott excitons is an exciting possibility using strong light matter coupling and making heterostructures of two or more excitonic materials. Such a heterostructure was realized using a combination of organic (J-aggregates) and inorganic ($WS_2$) excitonic material. Exciton absorption from $WS_2$ and J-aggregates are spectrally separated by 100 meV. However, when both are strongly coupled in the DBR microcavity it leads to observation of three polariton branches - UPB, LPB and MPB which contains, Frenkel, Wannier and Frenkel-Wannier mixed excitons, respectively. However, when the Stark effect is used to tune the lines, only LPB ($WS_2$) was tunable, while UPB (organic) absorption remained unaltered. By detuning the cavity length, the polariton's composition (Hopfield coefficients) in the MPB were tuned to be either Frenkel-rich ($\beta$) or Wannier-rich excitons ($\gamma$) as shown in Figure 6c, bottom panel. Rabi splitting was also shown to be tunable in $WS_2$ from 57 to 46 meV by applying an external voltage. Hence, the mixed exciton composition in MPB was modulated only by $WS_2$.(49) None the less the possibility to create and tune mixed excitons opens door to their exploration in other material combinations as well as device applications.

In the above section, we have discussed electrical control of hybrid exciton-photon states ranging from Frenkel excitons in SWCNTs, Wannier-Mott excitons in TMDCs, and Frenkel-Wannier-Mott hybrid excitons in a TMDC/organic heterostructures. TMDCs have thus far shown a pronounced effect in hybrid state modulation by electrical control. This is mainly because of their 2D nature with high degree of crystallinity. In the future, research efforts need to be focused not only on how to enhance coupling and electrical tuning but also on how to enhance radiative emission of the coupled states. Thus far, all discussed systems pertain to ensemble number of excitons and photons that couple with each other. In the following section we will focus our discussion on exciton-cavity interaction in the single-particle i.e. quantum limit achieved by confining and exciting only a few particles in the semiconductor medium.

## 7. Single photon emission (SPE) from multidimensional semiconductors

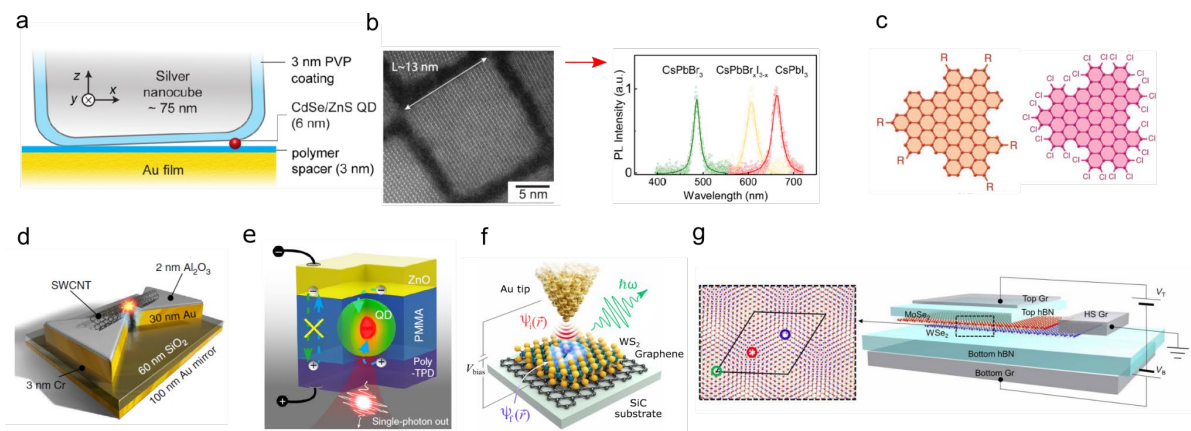

**Figure 7.** Excitonic single photon emitters. (a) II-IV (CdSe/ZnS QD) placed in plasmonic nanocavity. (b) Perovskite (PVSK) QDs observed under transmission electron microscope (b) and tunable PL by varying the halide ion composition (right panel in b). (c) Functionalized graphene QDs. (d) SWCNT suspended across a bowtie antenna. Schematic of electrically controlled SPE from (e) CdSe/CdS QDs, (f) tunneling electron induced emission from $WS_2$ defects and (g) Moiré heterobilayer ($MoSe_2$/$WSe_2$) with dual gates. Panel (a) - Adapted from ref.(213) Copyright 2015 American Chemical Society. Panel (b) - Reprinted with permission from ref.(214) Copyright 2015 American Chemical Society, Adapted from ref.(215) Copyright 2019 The American Association for the Advancement of Science (AAAS). Panel (c) - Reprinted with permission from ref.(216) 2018 Springer Nature CC BY. Panel (d) - Reprinted with permission from ref.(8) 2017 Springer Nature CC BY. Panel (e) -

When the two-level energy of the semiconductor materials integrated into a cavity remains unaltered due to the electromagnetic field inside the cavity, while the local density of optical states increases, the radiative recombination of the exciton increases significantly. This spontaneous emission with reduced exciton lifetime is called as Purcell effect. The Purcell effect while valuable across all domains and applications in optics and opto-electronics, is particularly critical and valuable in the quantum limit since it helps enhance the rate of emission of single photon emissions from quantized emitters. This is particularly critical as bright and pure single photon sources are paramount for any photons based quantum information applications. Here, we focus on advancements in II-IV QDs,(213) perovskite QDs (PVSK QDs),(214,215,220,221) graphene QDs (GQDs),(216) SWCNT(8,9,222)  and TMDC(223,224,233–235,225–232) as the quantum emitters and their integration with optical cavities to produce strong exciton-photon interactions. Specifically, we review their properties and understand how they compare with III-V QDs,(236–238) which are the workhorse of SPE. SPE performance such as second-order autocorrelation ($g^2(0)$), lifetime, and the operating temperature is pivotal to realize device integration. Low $g^2(0)$ and achieving room-temperature operation are the most sought properties in SPE candidates for device integration. In terms of emitter lifetime, two different applications are typically sought. Short exciton lifetime can be used for high-bit rate single photon source, while longer lifetime emitters are sought for use in quantum memory devices.

SPE from QDs is studied by either placing the emitters in a plasmonic cavity, diluting them e.g. for PVSK QDs  to a single-particle limit, functionalizing in the case of GQDs, and placing SWCNT on bowtie antennas as shown in Figure 7a-d.(213–216,220,221) In TMDC, excitons localized by inducing strain on the TMDC monolayer at the plasmonic particle edges or excitons trapped in Moiré lattice acts as SPE source. Further, electrical control of SPE from QDs, defects in TMDC and Moiré excitons are realized as shown in Figure 7e-g. We will first discuss the SPE from the plasmonic nanocavities and later present electrically controlled SPE, swiftly overcoming optical excitation challenges while realizing an electronic device.

II-IV QDs from CdSe/ZnS was the initial study demonstrating the Purcell effect to realize SPE in a plasmonic nanocavity (Figure 7a). Remarkable decrease (540-fold) in exciton lifetime and 1900-fold increase in the total emission intensity of the SPE from these II-IV QDs was reported.(213) PVSK QDs are another family of 0D materials where the halide ion composition can be easily tuned to vary the bandgap and emission emanating from them (Figure 7b). Nevertheless, the fundamentals of exciton dynamics in PVSK QDs are much more complicated than II-IV semiconductors. Rashba effect and exchange interaction induced exciton splitting into singlet and triplet states complicates the emission from PVSK QDs.(215) Due to these factors, PVSK QDs show a bright triplet state as the lower-lying state with accelerated radiative recombination (~100 ps),(239) while CdSe QDs are optically dark with a long lifetime (microseconds) at low temperatures. By drop casting weakly dispersed PVSK QDs on a glass substrate and using single photon spectroscopy, SPE was identified at 6 K(220) and at room temperature.(214) Single photon spectroscopy studies at 6 K, showed drastic decrease in the photoluminescence linewidth (~1 meV) with reduced blinking. Peculiar to QDs, additional decay channels such as non-radiative Auger recombination and radiative multiphoton recombination remains unavoidable. Although at low fluence (10 μW), the SPE is devoid of Auger recombination with a short lifetime (180-250 ps), at high fluence, multiphoton emission becomes dominant.(220) At room temperature, non-radiative Auger recombination was exploited to selectively suppress multiphoton emission to achieve high purity single photons under CW or pulsed excitation.(214) This is an interesting approach to quell additional decay channels and achieve pure SPE. SPE from QDs can also be extended to purely organic systems. Recently, chlorine-functionalized GQDs have shown SPE at room temperature with high photon purity (0.1) and long lifetimes (5 ns).(216) Given the flexibility in designing new molecules and adding functionalities to GQDs, SPE with wavelength tunability in the visible region will be of great interest to explore in this graphitic carbon system.

Besides 0D graphitic QDs, SPE is also possible from 1D nanotubes. Pristine 1D SWCNT when placed on plasmonic bowtie antennas have shown SPE along with quantum yield enhancement by 62% due to Purcell effect (Figure 7d).(8) Subsequently, functional groups covalently bonded to SWCNTs have led to $sp^3$ defect formation, which was shown to localize the excitons in an open cavity.(9,240,241) High photon purity (~0.01) with emission in the near-infrared region (1200-1500 nm) at room temperature from SWCNTs was demonstrated which are attractive properties for applications in the telecommunications domain.(9)

SPE from 2D TMDCs is observed only in monolayer due to their direct band gap nature. So far, SPE in TMDCs is realized in several ways, such as localized excitons,(225,226,234,235) QDs formation from TMDC,(227,242) electrical stimulus,(229,230,243) or deterministic strain by nanopillar in plasmonic nanocavity.(228,233) Out of all these approaches, strain-induced QD formation in TMDCs has been extensively studied for SPE. Using near-field spectroscopy, nanobubbles in $WS_2$/hBN have shown localized excitons with red-shifted emission (150 meV) compared to the primary excitons.(178) From prior studies, red-shifted emission was generally ascribed to SPE from localized excitons in WSe$_2$(225,226,234,235) and WSe$_2$ in plasmonic nanocavities.(224,231,244) Interestingly, the localized exciton dipole direction is out-of-plane compared to primary excitons (in-plane).(245) Similar observations about out-of-plane exciton dipole can explain the origin of SPE from strained induced TMDC QDs, which needs further exploration and verification.

Now we will discuss electrical control of SPE from excitonic materials. Electrically driven SPE was studied by sandwiching CdSe/CdS (without any cavity) between hole (poly-TPD) and electron (ZnO) conducting layer for electrical contact (Figure 7e).(217) Electrical excitation showed higher purity single photons (0.05) than optical excitation (0.08). Previously, we discussed that multiphoton emission becomes dominant at high fluence. Electrical injection suppresses these additional radiative biexciton formation, which is an indirect process. Fast recombination of trions inhibits biexciton (addition of hole/electron to trion) formation, thereby making single exciton recombination as dominant process for SPE. Also, increasing the thickness of the QD shell suppressed the charging effect during electrical excitation.(246) Collectively, room temperature electrical excitation of QDs in a device structure with high purity single photons is a significant leap for QD-based SPE device technology.

In contrast to SPE achieved by direct electrical excitation of QDs as discussed above, TMDC monolayer requires further manipulation to localize excitons by introducing defects or strain. Several strategies towards localizing excitons such as use of 2D/3D junctions of TMDCs/piezoelectric actuator to control strain,(223) controllably introducing atomic defects,(218) or forming interlayer excitons in Moiré potentials(219) have been demonstrated to achieve SPE in TMDCs as detailed below.

Piezoelectric actuators with $WSe_2$ monolayer on barium titanate were used to modulate strain by applying external voltage. By tuning voltage, the strain can be modulated from compressive to tensile up to 0.15% strain in the TMDC. Alternatively, electrons are injected into the defect energy levels in TMDCs using scanning tunneling microscopy (Figure 7f). These defects can be either sulfur vacancies ($Vac_S$) or chromium doped in W site ($Cr_W$). By applying electric bias, the emission emanating from the defect site showed single-electron to single photon conversion. Defect emission with high purity single photons and femtosecond lifetimes was observed. In another study, W replaced Se ($W_{Se}$) antisite defects in $WSe_2$/hBN have shown ultra-long (225 ns) lifetime.[232] From the above studies, it can be understood that tuning the right defect site in TMDCs makes them a promising candidate to explore either quantum emitter or quantum memory devices but more research is desired for deterministic placement of defects and identification of defect types based on the above two proposed applications. Recently, interlayer excitons trapped in Moiré potential formed by twisting TMDC heterobilayers by 60° have also shown SPE (Figure 7g). Interlayer excitons showed linear Zeeman splitting and Stark tunablity (40 meV) by applying magnetic and electric field, respectively. Long exciton lifetime and circular polarizability of interlayer excitons trapped in Moiré potential makes them different from strain induced TMDC SPEs.

So far, we have discussed SPE from all excitonic materials, and now we will compare their $g^2(0)$, exciton lifetime, and operating temperature, as shown in Figure 8. II-IV QDs in plasmonic nanocavities have demonstrated room-temperature operation with ps lifetime. Nevertheless, all QDs (II-IV, PVSK, and GQD) investigated in the open cavity have shown room temperature operation with long lifetimes (ns).

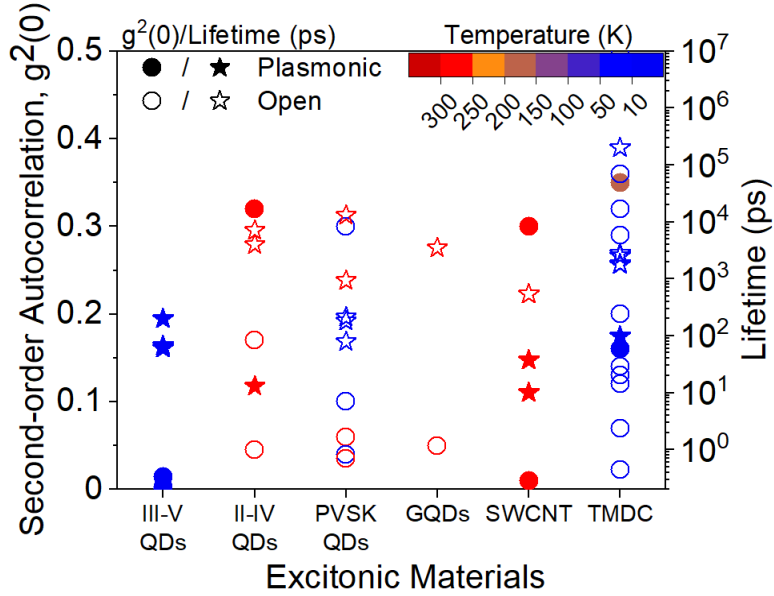

**Figure 8.** Figure of merit for single-photon emitters. Comparison of second-order autocorrelation function ($g^2(0)$) and lifetime for different excitonic materials - III-V QDs,(236–238) II-IV QDs,(213,217) perovskite QDs (PVSK QDs),(214,215,220,221,247) graphene QDs (GQDs),(216) single-walled carbon nanotube (SWCNT)(8,9,222) and TMDC(223,224,233–235,225–232) (a). The filled and open circles (or stars) denote $g^2(0)$ (or lifetime) for plasmonic nanocavity and open cavity, respectively. The color coordinate corresponds to the emitter's operating temperature as marked in the color scale.

Such properties may attract QDs for quantum memory devices. SWCNT emitting in telecommunication region with high purity and lifetime close to III-V QDs are a darkhorse, which holds great promise in next-generation communication devices. In contrast to QDs and SWCNTs, emission lifetime in TMDC SPE can be tuned by varying defect types in TMDCs or placing them in plasmonic nanocavities. Despite extensive work, the TMDC SPE mostly operates at low temperature (5 K) and shows a relatively large spread in $g^2(0)$, which makes them less viable as a reliable single-photon source. However, their ability to be integrated with other optical cavities and electrostatic tunability makes them attractive. To raise the operating temperatures and improve $g^2(0)$ confinement strategies such as use of Moiré potential in twisted bilayers and strain-induced confinement are some of the options to consider in future research investigations.

The above sections have discussed the fundamentals of light-matter coupling in excitonic semiconductor systems and further detailed the strength and weaknesses of various coupling modalities and semiconductors. In the next section, we will conclude this review with some of the emerging areas of research whose outcome strongly benefits from the light-matter coupling.

## 8. Emerging applications of coupled exciton-photons in solid state

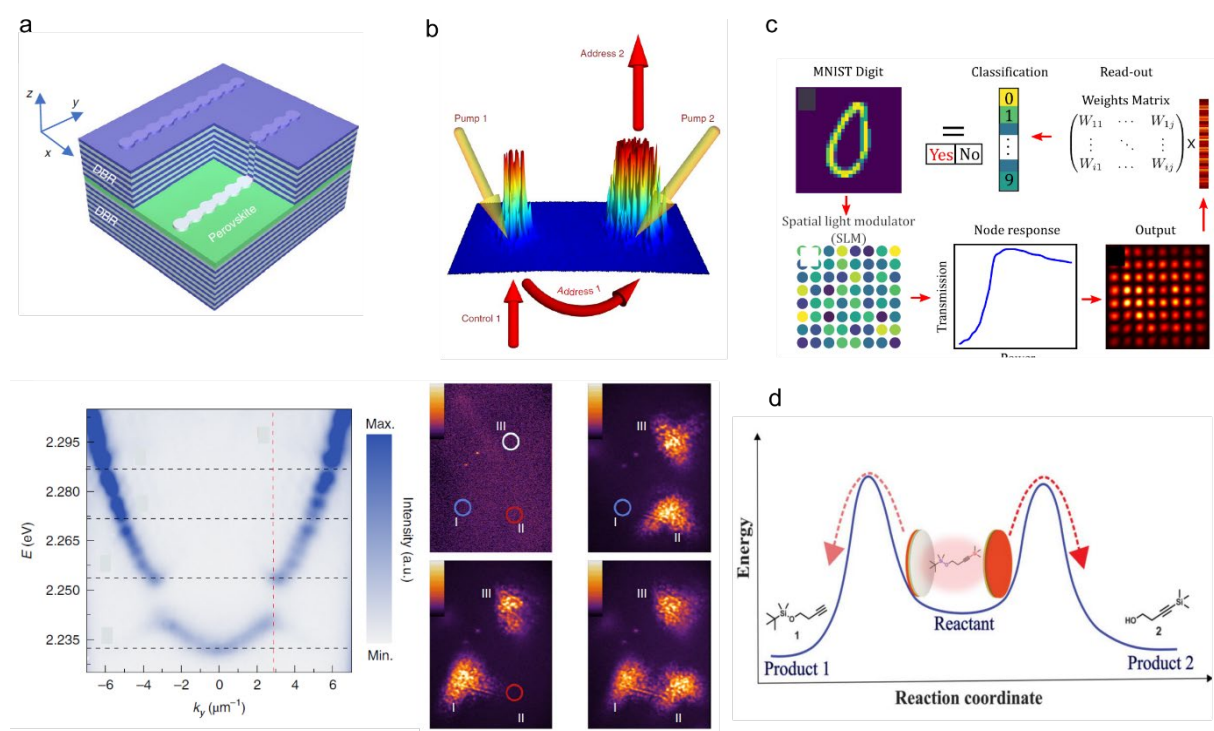

**Figure 9.** Emerging applications of polaritons. (a) Polariton lattice. (b) Organic optical switch. (c) Neuromorphic computing. (d) Polariton photochemistry. Panel (a) - Adapted from ref.(248) Copyright 2020 Springer Nature. Panel (b) - Adapted from ref.(65) Copyright 2020 Springer Nature. Panel (c) - Reprinted with permission from ref.(44) Copyright 2020 American Chemical Society. Panel (d) - Adapted from ref.(249) Copyright 2019 The American Association for the Advancement of Science (AAAS).

The electronic bandgap of a semiconductor is defined by its crystal structure and its composition. Changing the composition at atomic-scale will lead to tuning the electronic bandgap. However, using polariton states, the optical density of states in a semiconductor can be tuned based on the cavity design, polariton interaction, and the surrounding dielectric medium. This makes the concept of tuning optical density of states very powerful for investigations well beyond simple light-matter coupling for emission or absorption enhancement. The polariton lattice is one such concept of constructing an artificial lattice with different optical bandgaps using the optical properties from a single semiconductor. These lattices can be constructed in real space using electron-beam lithography,(248) ion beam milling(62) or laser patterning.(250) For instance, perovskites QDs can be filled into nanopillars etched in a DBR to form a perovskite lattice (Figure 9a). Using momentum and real space imaging, forbidden optical bandgaps (13.3 meV) emerging from the upper and lower polariton branches formed as a result of coupling excitons and photons in the real space lattice can be observed at room temperature (bottom panel, Figure 9a). Similarly artificial lattices have been extended to other symmetries such as honeycomb structure (hexagonal-symmetry), Lieb lattice (square-symmetry).(251) Real space lattice engineering to tune coupling and dispersions of hybrid exciton-photon states opens new avenues ranging from realizing flat band Hamiltonians with spin-orbit coupling(252) to topological polariton devices,(253) and even optical simulators(254) for interacting systems in a non-linear regime.

Another tantalizing possibility arising from engineering dispersions of coupled light-matter excitations is to use the frictionless flow of polaritons is exploiting them in optical switches.(64,65) Polariton states from organic materials strongly coupled to external DBR cavity were used to demonstrate an optical switch by cascade amplification. For this purpose, excitons dressed with vibrational energy forming "hot excitons" are required to exploit vibron-mediated energy relaxation.(65) The rigidity in ladder-type polymers offers such high-energy vibron mode with high oscillator strength. Upon forming polariton states with a DBR cavity, two pumps are used to achieve cascaded amplification in two-stages as shown in Figure 9b. It can be noticed that the condensate emission achieved from the pump 1 is further amplified by pump 2. This was used to demonstrate "OR" binary logic gate as shown in the bottom panel, Figure 9b. Furthermore, the non-linearity of polaritons can be used in complex pattern recognition which underlies the basis of most neuromorphic computing platforms. To create an artificial non-linear, optical neuron III-V semiconductor quantum wells forming polaritons in a DBR cavity at low temperatures were used.(44) Hand-written digits were converted into a matrix that is readable by the nodes in the polariton network. Using spatial light modulator, the digits were encoded into a laser pulse with controllable intensity and phase. Passing this digit-encoded laser into the polariton cavity detected the digits in the transmitted pattern with outstanding accuracy as compared to linear classifiers.

Finally, hybrid states can also be used to modify chemical reactions by embedding the reactants in a cavity (Figure 9d).(249) Vibrational strong coupling was achieved by vibrational modes from organics such as silane (liquid reactant) resonantly interacting with the cavity modes. Two different products can be formed based on silane cleavage occurring through Si-C or Si-O bond. Using infra-red spectroscopy, the products emerging under strong coupling were detected with high accuracy. Strong splitting of infra-red peaks corresponding to Si-C or Si-O bond was observed by tuning the cavity mode. This technique can be an additional experimental knob to select and accelerate the desired chemical reaction, thereby tilting the chemical landscape using the light-matter coupling phenomenon and opening a fundamentally novel direction in light-driven catalytic chemistry.

## 9. Conclusion and Outlook

In this article, we have reviewed the role of exciton-photon interactions in low-dimensional materials systems ranging from their basic science to measurement to applications. Properties and features of hybrid exciton-polariton states from strong coupling of ensemble particles and Purcell enhanced emission of single-particle excitations from weak coupling are delineated for these materials.

Over the past two-decades or so tremendous progress has been achieved by the scientific community in not only producing and isolating novel excitonic materials but also developing measurement and fabrication techniques to resonantly couple photons into these materials. While the use of dielectric Bragg mirrors and plasmonic antennas are more mainstream, the use of extreme dielectric constants of the excitonic materials themselves to induce lossy Fabry-Perot like modes in sub-wavelength dimension semiconductors is a new and interesting avenue for further exploration. Given the simplicity of the concept and geometry of resulting samples a lot more activity is expected in this area in the future including in the space of photochemistry and photocatalysis.

Another interesting observation that is evident from the literature is the relative scarcity of studying exciton-photon interactions with electrical injection of carriers to produce photons directly inside the cavity. There are many possible reasons for this including lack of appropriate, low-resistance contacting schemes to incompatibility of certain excitonic materials to micro/nano fabrication processes. However, this area, does need more attention and development since electrical injection is the likely path to wide-spread, practical exciton-photonic devices ranging from low-threshold polaritonic lasers to electro-optic modulators and switches.

Further, given the maturity of the field of quantum optics and quantum information theory, it is imperative that excitonic materials will be looked at as possible candidates for miniature, integrated, solid-state quantum light sources. In this regard, III-V QDs are the current standard bearers but the lack of emission in telecom wavelengths is a likely impediment. The key requirements for ideal quantum photon source are many-fold ranging from high brightness to tunability and ease of integration to deterministic positioning, low $g^2(0)$ and alignment with telecom spectrum. Meeting all of these in one material system is challenging and likely to remain one for some time. Towards that end, 1D excitonic materials like carbon nanotubes and even defect/strain induced quantum emitters in 2D materials are viable options and must therefore be investigated with more vigor and creative ideas particularly with regards to scaling up, adjusting emission wavelength in telecom range and deterministic positioning.

Finally, in terms of techniques and sample/materials development there are several major points worth nothing and acting upon. First of all, given the short life-time of excitons in most discussed semiconductors it is imperative to develop near and field, real-space transient imaging techniques for coupled exciton-photons. While such techniques already exist in the far-field for excitons, time-resolved near-field techniques for excitons that normally lie in visible part of the spectrum are rare. Integrating time- and space-resolved optical measurements in one-setup will advance our understanding of exciton properties and new functionalities at the nanoscale. Replacing the laser source from continuous-wave to pulsed laser to acquire time-resolved photoluminescence measurements in the near-field spectroscopy setup can be a first step in this regard. For instance, a combination of near-field and time-resolved measurements investigated in polymer blends were able to identify the competing process between energy transfer routes and trap states.(255–257) Embarking in this research direction will open new avenues for quantum optics, nanophotonics, and metamaterials research community. Further, nonlinear optical processes such as two-photon absorption and second harmonic generation will improve nanoscale imaging resolution by selecting only the optical ranges of interest. Beyond optical response in the visible range, additional information from the material can be identified from the infra-red region using near-field technique. Recently, infra-red vibrational spectroscopic studies have shown label-free chemical information and cation vibrational dynamics in liquids(258) and perovskites(259), respectively. Extending the laser excitation up to UV range in near-field spectroscopy studies will open up a vast amount of materials such as perovskite QDs and its superlattice,(260) organics on 2D hybrids(261) and biological materials, which have shown extraordinary optical properties in the far-field. Identifying tip materials and modified optics, which can resonate at UV range, will also be critical to realizing this step.

Overall, the past two decades have witnessed tremendous progress in excitonic materials and optical techniques to couple photons strongly with them and characterize them. The coming decades are expected to take this field which has largely been in the domain to basic science into a domain to practical technology with improved materials, interfaces and optical cavities. Likewise, on the basic science front it is expected that improvement in imaging and spectroscopic techniques will shine more light on the transient nature and dynamics of coupled exciton-photons and open new avenues for fundamental research in both condensed matter physics, chemistry and quantum information sciences.

## Acknowledgment

D.J. and K.J. acknowledge primary support for this work by the U.S. Army Research Office under contract number W911NF-19-1-0109. S.B.A gratefully acknowledges the funding received from the Swiss National Science Foundation (SNSF) under the Early Postdoc Mobility program (grant 187977) to conduct this work.

**TOC**

Exciton-photon coupling leading to novel light-matter interactions in low-dimensional semiconductors in the strong and weak coupling limit is reviewed. Examples of novel photonic and opto-electronic phenomena in both near and far-field are discussed. Devices applications and future outlook for the field are discussed.

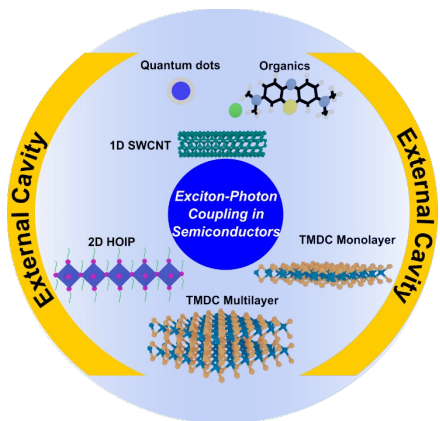